\documentclass[twocolumn]{aastex61}

\usepackage{apjfonts,amsmath,color,textcomp,url,graphicx,subfigure}
\usepackage{soul}

\newcommand{\kms}{\hbox{km\,s$^{-1}$}}

\newcommand{\Mjup}{$M_{\mathrm{Jup}}$}

\newcommand{\masyr}{$\mathrm{mas}\,\mathrm{yr}^{-1}$}
\newcommand{\teff}{$T_{\rm eff}$}
\newcommand{\besancon}{Besan\c con}
\newcommand{\objashort}{2M2250+2325}
\newcommand{\objbshort}{2M1219+0154}
\newcommand{\objalong}{2MASS~J22501512+2325342}
\newcommand{\objblong}{2MASS~J12193316+0154268}

\accepted{ApJ}

\shorttitle{The discovery of 2M2250+2325~B}
\shortauthors{Desrochers et al.}

\begin{document}

\title{BANYAN. X. DISCOVERY OF A WIDE, LOW-GRAVITY L-TYPE COMPANION TO A FAST-ROTATING M3 DWARF}

\author[0000-0003-4142-905X]{Marie-Eve Desrochers}
\thanks{Based on observations obtained at the Canada-France-Hawaii Telescope (CFHT) which is operated by the National Research Council of Canada, the Institut National des Sciences de l''Univers of the Centre National de la Recherche Scientique of France, and the University of Hawaii.}
\affil{Institut de Recherche sur les Exoplan\`etes, Universit\'e de Montr\'eal, D\'epartement de Physique, C.P.~6128 Succ. Centre-ville, Montr\'eal, QC H3C~3J7, Canada}
\email{desrochers@astro.umontreal.ca}

\author[0000-0003-3506-5667]{\'Etienne Artigau}
\affil{Institut de Recherche sur les Exoplan\`etes, Universit\'e de Montr\'eal, D\'epartement de Physique, C.P.~6128 Succ. Centre-ville, Montr\'eal, QC H3C~3J7, Canada}
\author[0000-0002-2592-9612]{Jonathan Gagn\'e}
\affiliation{Carnegie Institution of Washington DTM, 5241 Broad Branch Road NW, Washington, DC~20015, USA}
\affiliation{NASA Sagan Fellow}
\author[0000-0001-5485-4675]{Ren\'e Doyon}
\affil{Institut de Recherche sur les Exoplan\`etes, Universit\'e de Montr\'eal, D\'epartement de Physique, C.P.~6128 Succ. Centre-ville, Montr\'eal, QC H3C~3J7, Canada}
\author[0000-0002-8786-8499]{Lison Malo}
\affil{Institut de Recherche sur les Exoplan\`etes, Universit\'e de Montr\'eal, D\'epartement de Physique, C.P.~6128 Succ. Centre-ville, Montr\'eal, QC H3C~3J7, Canada}
\author[0000-0001-6251-0573]{Jacqueline K. Faherty}
\affiliation{Department of Astrophysics, American Museum of Natural History, Central Park West at 79th St., New York, NY 10024, USA}
\author[0000-0002-6780-4252]{David Lafreni\` ere}
\affil{Institut de Recherche sur les Exoplan\`etes, Universit\'e de Montr\'eal, D\'epartement de Physique, C.P.~6128 Succ. Centre-ville, Montr\'eal, QC H3C~3J7, Canada}

\begin{abstract}

We report the discovery of a substellar-mass co-moving companion to 2MASS~J22501512+2325342, an M3 candidate member of the young (130--200\,Myr) AB~Doradus Moving Group (ABDMG). This L3\,$\beta$ companion was discovered in a 2MASS search for companions at separations of 3--18\arcsec\ from a list of 2\,812 stars suspected to be young ($\lesssim$\,500\,Myr) in the literature, and was confirmed with follow-up astrometry and spectroscopy. Evolutionary models yield a companion mass of $30_{-4}^{+11}$\,\Mjup\ at the age of ABDMG. The 2MASS~J22501512+2325342~AB system appears to be a spatial outlier to the bulk of ABDMG members, similarly to the young 2MASS~J22362452+4751425~AB system. Future searches for young objects around these two systems would make it possible to determine whether they are part of a spatial extension of the known ABDMG distribution.

\end{abstract}


\section{INTRODUCTION}\label{sec:intro}

Direct imaging is a unique exoplanet detection method, as it makes it possible to constrain the properties of exoplanets directly from their emitted light. The large contrast ratio between Sun-like stars and self-luminous young planets however poses a significant challenge to their detection. As demonstrated by recent discoveries, this challenge can be partially alleviated by searching for distant companions (1--20\arcsec) around late-type, faint stars. This strategy provides a favorable contrast ratio, allowing for the detection of lower-mass companions. In the best-case scenarios, self-luminous planetary and substellar companions can be detected with seeing-limited observations. Almost all planetary-mass companions that were directly imaged to date are younger than 200\,Myr old \citep{2016PASP..128j2001B} and relatively well-separated from their host star ($>$\,50~AU). Substellar companions can be found around Gyr-old stars, but probing younger systems is a good strategy to achieve lower-mass detections at a fixed contrast ratio. Some companions, such as 2MASS~J02192210--3925225~B \citep{2015ApJ...806..254A} and VHS~J125601.92--125723.9~b \citep{2015ApJ...804...96G}, were detected in 2MASS images after subtracting the central point-spread function (PSF) of the host star. These recent detections at the brown dwarf to planetary-mass limit based on 2MASS images suggests that a more thorough search for similar objects may yield new discoveries.

Here we report the results of such a survey, in which we discovered a low-gravity brown dwarf companion to the M3 AB~Doradus candidate member \objalong\ (\objashort), and a likely field M9 companion to the M3 dwarf \objblong\ (\objbshort) that was previously identified as an ABDMG candidate member. A description of the sample of young stars that were searched for companions is presented in Section~\ref{sample}. The method used to identify candidates is described in Section~\ref{sstrategy}, followed in Section~\ref{follow-up} by a detailed description of the follow-up observations that allowed us to verify the substellar nature of the companion candidates. The results and analysis are described in Section~\ref{sec:Results}. In Section~\ref{sec:member}, the young moving group membership of \objashort\ and \objbshort\ are discussed. The confidence and efficiency of the detection method and the fundamental properties of the newly discovered system are discussed in Section~\ref{discussion}. This work is concluded in Section~\ref{conclusion}.

\section{SURVEY SAMPLE}
\label{sample}

A list of all plausibly young ($\lesssim$\,500\,Myr) stars within 100\,pc of the Sun was compiled from the literature, mainly from members or candidate members of nearby young associations (e.g., see Gagn\'e et al., submitted to ApJS). Objects with other youth indicators such as strong H$\alpha$ emission, low gravity, lithium absorption, chromospheric activity or X-ray emission were also included in the sample. The resulting sample is not complete by construction, as it relies on a multitude of surveys with differing selection criteria, and should therefore not be used for statistical population studies.

From the initial list of 4\,303 potentially young stars, only those outside of the Galactic plane ($|b| > 15$\textdegree) were selected to avoid confusion in the cross-matching of catalogs in crowded fields from blends of stars and only those fainter than $J = 8$ were selected to limit for favorable contrast and standardizes the sample. The resulting list of 2\,812 stars is mostly constituted of K and M dwarfs. At a typical distance of $\sim$\,40\,pc, the faintness criterion $J > 8$ selects stars later than $\sim$\,K4 \citep{1982PASJ...34..529M}.

The young stars in our sample that are known members or candidate members of 20 known young associations and clusters are listed in Table 1. The reference column lists literature work that is relevant to these associations and clusters. Their potential membership makes it possible to constrain their ages and their distances (based on sky position and proper motion), making them particularly valuable targets.

\section{IDENTIFICATION OF CANDIDATE COMPANIONS}
\label{sstrategy}

The search strategy was designed to recover known red companions similar to 2MASS~J02192210--3925225~B \citep{2015ApJ...806..254A} and VHS~J125601.92--125723.9~b \citep{2015ApJ...804...96G}. In each band of the original 2MASS Atlas images ($J$, $H$ and $K_S$), The PSF of each young star in the final sample was correlated with those of all other stars in the initial 2\,812-stars sample to select the 20 most similar PSFs over a 36\arcsec-wide square box, and those were median-combined to create a reference PSF. These reference PSFs were subtracted from the 2MASS frames of each young sample star, and point sources at separations of 3--18\arcsec\ from the young star were compiled. The lower limit of 3\arcsec\ was chosen based on the 2MASS spatial resolution ($\sim$\,2.5--3\arcsec) which is a consequence of pixel sampling ($\sim$\,2$''$) and atmospheric seeing ($\sim$\,1--1.5\arcsec). The upper limit of 18\arcsec\ was chosen to minimize the number of contaminating background sources, and corresponds to physical separations of 100--800\,AU at distances of $\sim$\,5--45\,pc, which is typical for nearby binaries and high-mass ratio stellar-substellar systems \citep{2016PASP..128j2001B}.

The compiled point sources were defined as candidates only if they: (1) were detected above 2$\sigma$ in $J$ band, and above 3$\sigma$ in both $H$ and $K_S$ bands; (2) had $J-K_S$ in the range 1--3; and (3) were not detected in the DSS2-red survey. The constraint on the $J$-band detection is relaxed to ensure detection of red $J-K_S$ objects that are typical of late spectral type ($\gtrsim$\,L0). The DSS2-red survey is limited to magnitudes of $r < 20.5$\,mag \citep{djorgovski2013sky}, and the optical non-detection constraint thus translates to $r > 17$\,mag at $\approx$\,50\,pc, which corresponds to masses of $\lesssim$\,60\,\Mjup\ at $\sim$\,150\,Myr, or $\lesssim$\,20\,\Mjup\ at 20\,Myr \footnote{See BT-Settl SDSS-band isochrones at \url{https://phoenix.ens-lyon.fr/Grids/BT-Settl/CIFIST2011\_2015/}} \citep{2012RSPTA.370.2765A}.

The 30 candidates that resulted from these selection criteria were visually inspected to eliminate PSF subtraction residuals, instrumental optical ghosts or elongated sources. When near-infrared coverage deeper than 2MASS such as UKIDSS \citep{2007MNRAS.379.1599L}, SDSS $i$- or $z$-band \citep{2015ApJS..219...12A}, or the SIMP survey \citep{2016ApJ...830..144R} was available, candidates not detected in the corresponding band of the deeper survey were eliminated, and those detected were required to be located at the same separation and position angle.

Of the seven candidates selected by these additional criteria, five are already listed as substellar companions in the literature  ; 2MASS~J0219--3925~B \citep{2015ApJ...806..254A}, LP~261--75~B (\citealt{2000AJ....120..447K} ; \citealt{2006PASP..118..671R}; \citealt{2004AJ....127.2948V}), G~196--3 B (\citealt{1998Sci...282.1309R}; \citealt{1538-3881-137-2-3345}; \citealt{2014ApJ...783..121G}; \citealt{2014AA...572A..67Z}; \citealt{2014AA...572A..67Z}), 2MASS~J1256--1257~b \citep{2015ApJ...804...96G}, and 2MASS~J2322--6151~B (\citealt{2008AJ....136.1290R}, \citealt{2016ApJS..225...10F}, \citealt{2015ApJS..219...33G}). All known companions with 3--18\arcsec\ separations that respect the selection criteria described above were recovered in our survey. Companions such as 2MASS~J12073346-3932539~b (0.78\arcsec; \citealt{2004AA...425L..29C}) were not recovered due to their small angular separation, and others such as 2MASS~J22362452+4751425~b ($J \sim$\,15.8; \citealt{2017AJ....153...18B}) were not recovered due to their faintness.

\begin{figure*}
    \centering
    \subfigure[\objbshort]{\includegraphics[width=0.488\textwidth, clip=true, trim= 0cm 2cm 0cm 2cm]{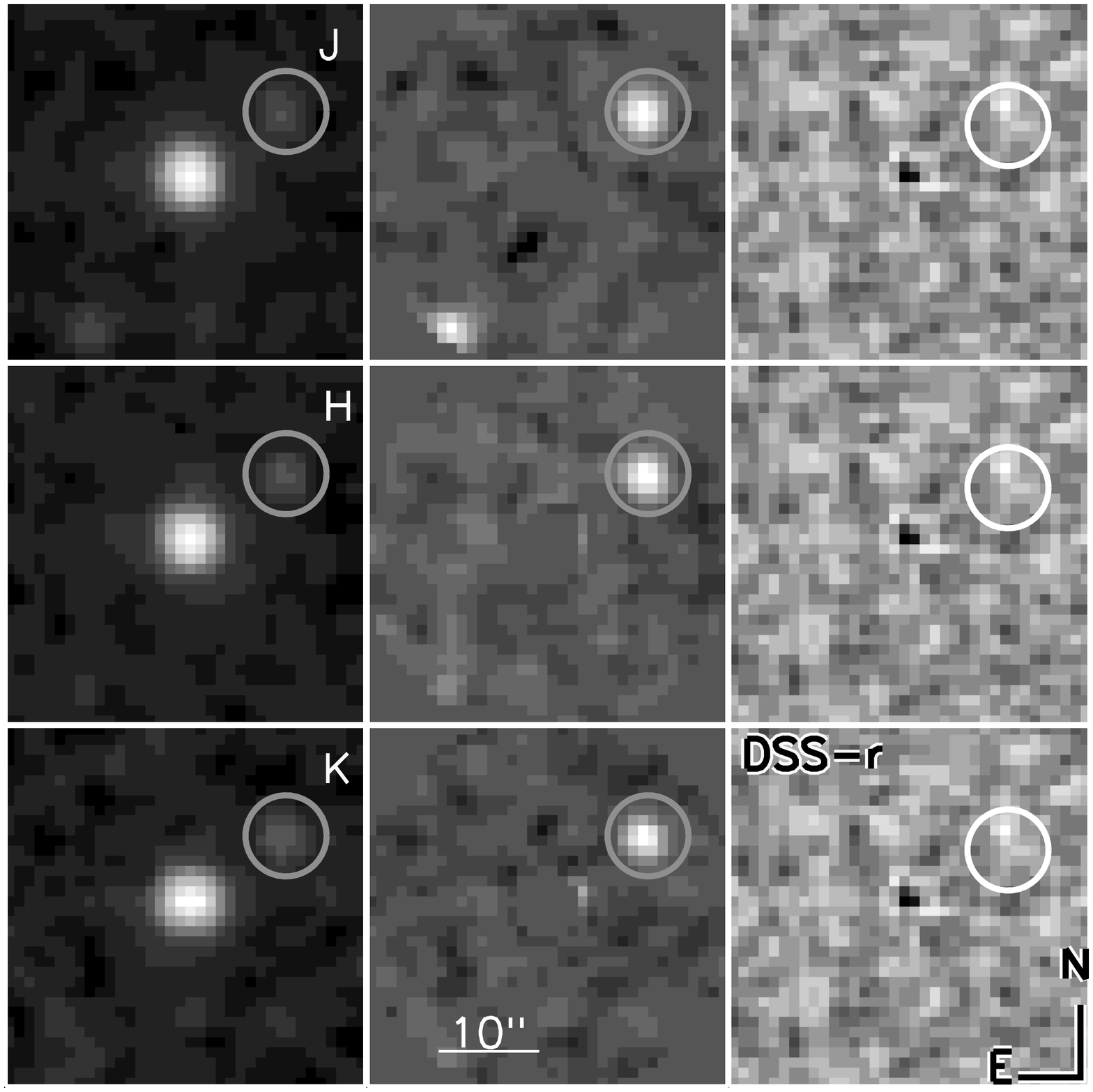}}\label{psf1219}
    \subfigure[\objashort]{\includegraphics[width=0.488\textwidth, clip=true, trim= 0cm 2cm 0cm 2cm]{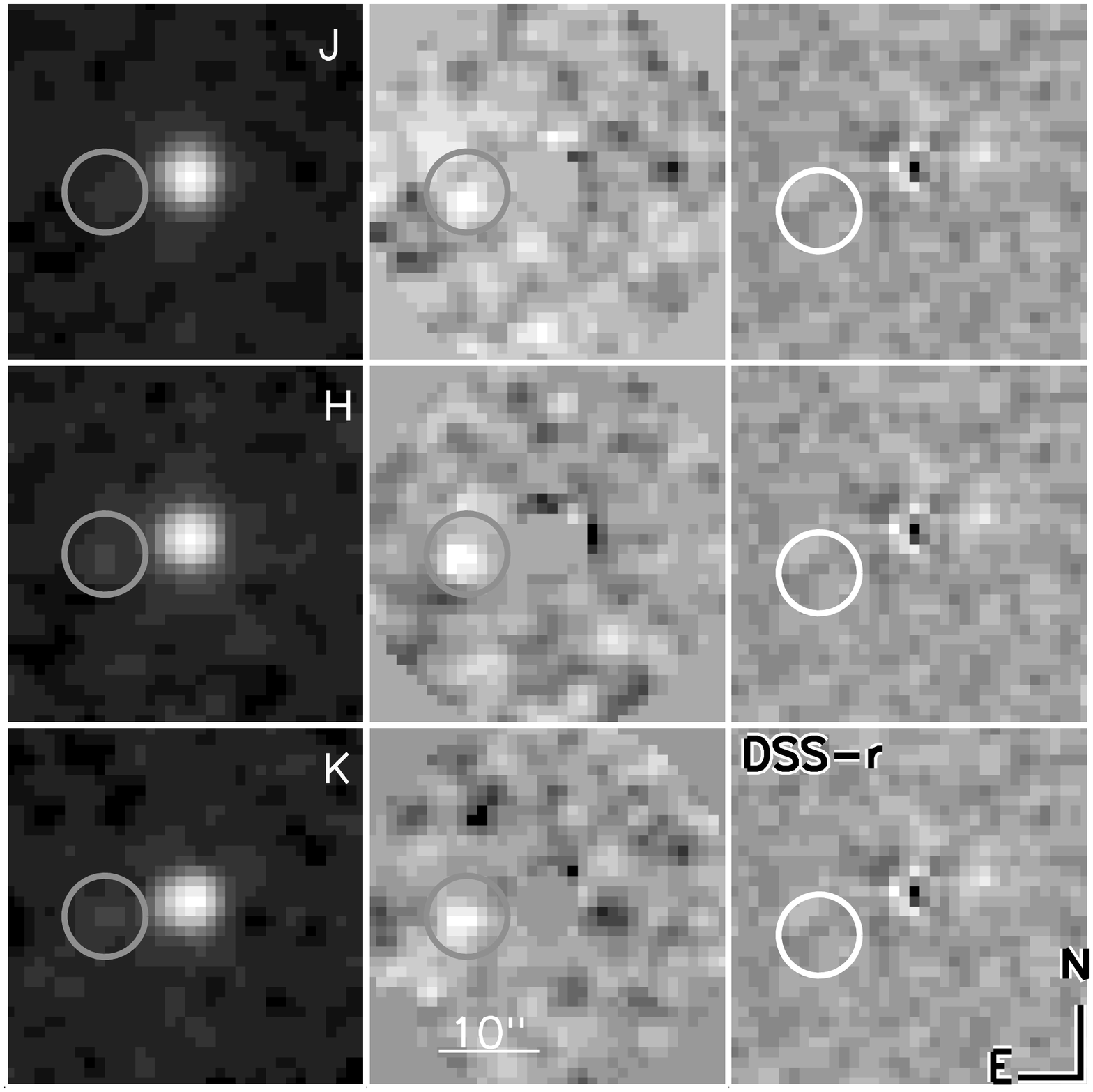}}\label{psf2250}
    \caption{2MASS and DSS-$r$ 36\arcsec$\times$36\arcsec\ frames around \objbshort and \objashort. The original 2MASS frames and the resulting signal-to-noise maps from the PSF-subtracted 2MASS frames are displayed in the first two columns, and the three rows correspond to the $J$, $H$ and $K_S$ bands. The third column shows the PSF-subtracted DSS-$r$ frame. The white circles outline the position of the companion in the 2MASS residual images.}
    \label{PSFsubstraction}
\end{figure*}

The two remaining candidates, \objblong~B and \objalong~B  (hereafter: \objbshort~B and \objashort~B), are new discoveries. The 2MASS and DSS-$r$ detection images of both new candidates are presented in Figure~\ref{psf1219}.

\begin{deluxetable}{lchl}
\tablecaption{Sources for the young star sample \label{refsample}}
\tablehead{\colhead{Association} & \colhead{Num. Stars} & & \colhead{Ref.}}
\startdata
TW~Hya & 107 & 5--15 & 1,2\\
$\beta$~Pictoris & 141 & 20--26 & 1,3,4 \\
Tucana-Horologium (THA) & 295 & 20--40 & 1,5 \\
Columba & 127 & 20--40 & 1,6 \\
Carina (CAR) & 50 & 20--40 & 1,6 \\
Argus/IC~2391 & 80 & 30--50 & 7 \\
AB~Doradus & 230 & 110--130\tablenotemark{a} & 1,8,9 \\
Carina-Near & 1 & $\sim$ 200 & 10 \\
$\epsilon$~Cha & 4 & $\sim$ 10 & 11 \\
Ursa Major& 14 & $\sim$ 500 & 11,12 \\
Hercules-Lyrae & 10 & $\sim 250$ or Stream & 13,14 \\
$\eta$~Cha & 12 & $\sim$ 8 & 1,11 \\
Octans & 50 & 30--40 & 5,15 \\
Octans-Near & 1 & 30--100 & 16 \\
Hyades Cluster & 424 & 625-650 & 11 \\
Upper~Scorpius & 99 & 12--14 & 17 \\
Chamaeleon~I & 5 & 1--2 & 18 \\
$\sigma$~Orionis & 4 & $\sim$ 3 & 19 \\
Taurus & 9 & 1-2 & 20 \\
Pleiades & 7 & $\sim$ 100 & 21 \\
Field/Ambiguous & 1142 & 20-500 & 21--26\\
\enddata
\tablerefs{(1)~\citealt{2015MNRAS.454..593B}, (2)~\citealt{2013ApJ...762..118W}, (3)~\citealt{2014ApJ...792...37M}, (4)~\citealt{2014MNRAS.438L..11B}, (5)~\citealt{2014AJ....147..146K}, (6)~\citealt{2008hsf2.book..757T}, (7)~\citealt{2013MNRAS.431.1005D},
(8)~\citealt{2005ApJ...628L..69L}, (9)~\citealt{2013ApJ...766....6B},
(10)~\citealt{2006ApJ...649L.115Z}, (11)~\citealt{2004ARAA..42..685Z},
(12)~\citealt{2003AJ....125.1980K}, (13)~\citealt{2013AA...556A..53E},
(14)~\citealt{2014ApJ...786....1B}, (15)~\citealt{2015MNRAS.447.1267M},
(16)~\citealt{2013ApJ...778....5Z}, (17)~\citealt{2012ApJ...746..154P},
(18)~\citealt{2007ApJ...662..413K}, (19)~\citealt{2007ApJ...662.1067H},
(20)~\citealt{2008ApJ...686L.111K}, (21)~\citealt{2008AA...483..453F}, (22)~\citealt{2013AJ....146..154M}, (23)~\citealt{2012AJ....143..135V},
(24)~\citealt{2009ApJ...699..649S}, (25)~\citealt{2005AA...430..165F},
(26)~\citealt{1995AJ....110.2862E}}
\end{deluxetable}

\section{OBSERVATIONS}
\label{follow-up}

Additional near-infrared imaging was obtained to better constrain the colors of the companions and confirm their common proper motion with the primary stars, and spectroscopic observations of both primaries and companions were obtained to constrain their physical properties.

\subsection{CPAPIR Imaging}

$J$, $H$ and $K$-band images were obtained with the CPAPIR infrared camera \citep{2006PhDT........18A} at the Mont-M\'egantic Observatory on 2015 May 26 (\objbshort\ and \objashort) and 2016 May 20 (\objashort ; see Figure~\ref{image2250}). CPAPIR has a 30\arcmin$\times$ 30\arcmin\ field of view which makes it possible to use numerous field stars as photometric calibrators. Observations were obtained under a 2\arcsec\ seeing and photometric conditions on both nights. A random 2$\arcmin$  dither pattern of 30 images was used for a total of 600\,s of integration per bandpass. The sky subtraction was performed using a median sky frame constructed from the dataset itself. Flat-fielding of images was done with calibration frames obtained at the end of the by observing an illuminated flat screen in the dome.

The astrometric calibration of frames was performed by anchoring the solution to the GAIA~DR1 \citep{2016yCat.1337....0G} on individual frames prior to combination. The photometry measurements were performed through aperture photometry. The zero points were defined by using all 2MASS stars in the Point Source Catalog within 8$\arcmin$  of the targets and their uncertainties were determined from the dispersion of the differences to the median divided by the square root of the number of calibrators. 

\subsection{Spectroscopy}

\subsubsection{ESPaDOnS at CFHT}
High-resolution optical spectra were obtained with the `star+sky' mode of ESPaDOnS at the Canada France Hawaii Telescope for \objbshort~A and \objashort~A on 2015 July 5 and 2015 July 1, respectively through program number 15AC24. These data are useful to measure the radial velocities and verify moving group membership, and they also allow to measure projected rotational velocities and lithium absorption, both of which are useful indications of youth.

The observations yielded a resolution of $\lambda/\Delta\lambda \sim$ 68\,000 across 370--1\,050\,nm. Total exposure times of 1200\,s were obtained for each star, resulting in a signal-to-noise ratio of $\sim$\,70 at 809\,nm. Two slowly rotating radial velocity standards were observed $\sim$ 30 and $\sim$ 15 \,min before \objbshort~A to ensure a stable wavelength calibration. The data were reduced by the CFHT pipeline UPENA (Version 2.12; 2006 Apr. 20), based on the Libre-ESpRIT software \citep{1997MNRAS.291..658D}\footnote{A description of the pipeline is availible at \url{http://www.cfht.hawaii.edu/Instruments/Upena/}.}. 

\subsubsection{FIRE at Magellan/Baade}
Near-infrared spectra were obtained for \objbshort~A and \objbshort~B on 2015 May 31 with the Folded-port InfraRed Echellette (FIRE; \citealp{2008SPIE.7014E..0US,2013PASP..125..270S}) at the Magellan/Baade telescope to measure their radial velocities and verify their moving group membership. Obtaining a radial velocity for both components is also useful to strengthen their co-moving nature, and medium-resolution spectroscopy is also useful to investigate for youth using gravity-sensitive spectral indices. The 0.6\arcsec\ slit was used, yielding a resolving power of $R \sim$\,6\,000 across 0.8--2.5\,$\mu$m. Six 90\,s exposures were obtained for \objbshort~A, and four 900\,s exposures were obtained for \objbshort~B in repeated ABBA patterns along the slit, resulting in respective signal-to-noise ratios of $\sim$\,500 and $\sim$\,140 per pixel at $\sim$\,1.6\,$\mu$m.

Ten 1\,s and ten 3\,s internal flat field exposures were obtained at the beginning of the night to properly illuminate all red and blue orders, and were used to correct pixel response of the detector. Ten 1\,s dome flat exposures were also obtained at the beginning of the night to obtain the slit illumination function. The A0-type star HD~105782 was observed 80\,min after the science targets at an airmass difference of 0.14, using six 90\,s exposures to achieve telluric and instrument response corrections. One 10\,s internal Th--Ar lamp exposure was obtained immediately after both the science and telluric sequences to obtain a wavelength solution at each telescope pointing position. The data were reduced using the Interactive Data Language (IDL) Firehose~v2.0 package (\citealp{2009PASP..121.1409B,zenodofirehose}\footnote{Available at \url{https://github.com/jgagneastro/FireHose\_v2/tree/v2.0}}; see \citealp{2015ApJS..219...33G} for more details on this reduction package).

\subsubsection{SpeX at IRTF}
A low-resolution near-infrared spectrum was obtained for \objashort~B with SpeX \citep{2003PASP..115..362R} at the InfraRed Telescope Facility (IRTF) on 2015 August 21 to verify its substellar nature and measure gravity-sensitive features. The 0.8\arcsec\ slit was used with the prism mode, yielding a resolving power of $R \sim 75$ in the 0.8--2.5\,$\mu$m range. Twenty one exposures of 150\,s each were obtained in repeated ABBA patterns along the slit, yielding a signal-to-noise ratio per pixel of $\sim$\,60 at $\sim$\,1.6\,$\mu$m. A standard calibration sequence consisting of five 0.5\,s internal flat field exposures and one 0.5\,s internal Ar lamp exposure were obtained in between the science and telluric sequences, to correct for relative pixel response variations and achieve wavelength calibration, respectively. The A0-type standard HD~215690 was observed immediately after the science target (using twenty one 0.5\,s exposures) and at a similar airmass to correct for telluric absorption and instrumental response. The data were reduced with the spextool v4.0 beta IDL package \citep{2004PASP..116..362C}, and telluric corrections were achieved using the xtellcor IDL routine \citep{2003PASP..115..389V}.

\section{RESULTS AND ANALYSIS}
\label{sec:Results}

This section reports on the spectral and kinematic properties of each component of the two new co-moving systems discovered here. All measurements are compiled in Table~\ref{alll}.

\subsection{Host Star Spectral Properties}

\begin{figure*}
    \centering
    \subfigure[\objbshort~A]{\includegraphics[width=0.488\textwidth, clip=true, trim= 0cm 5cm 0cm 5cm]{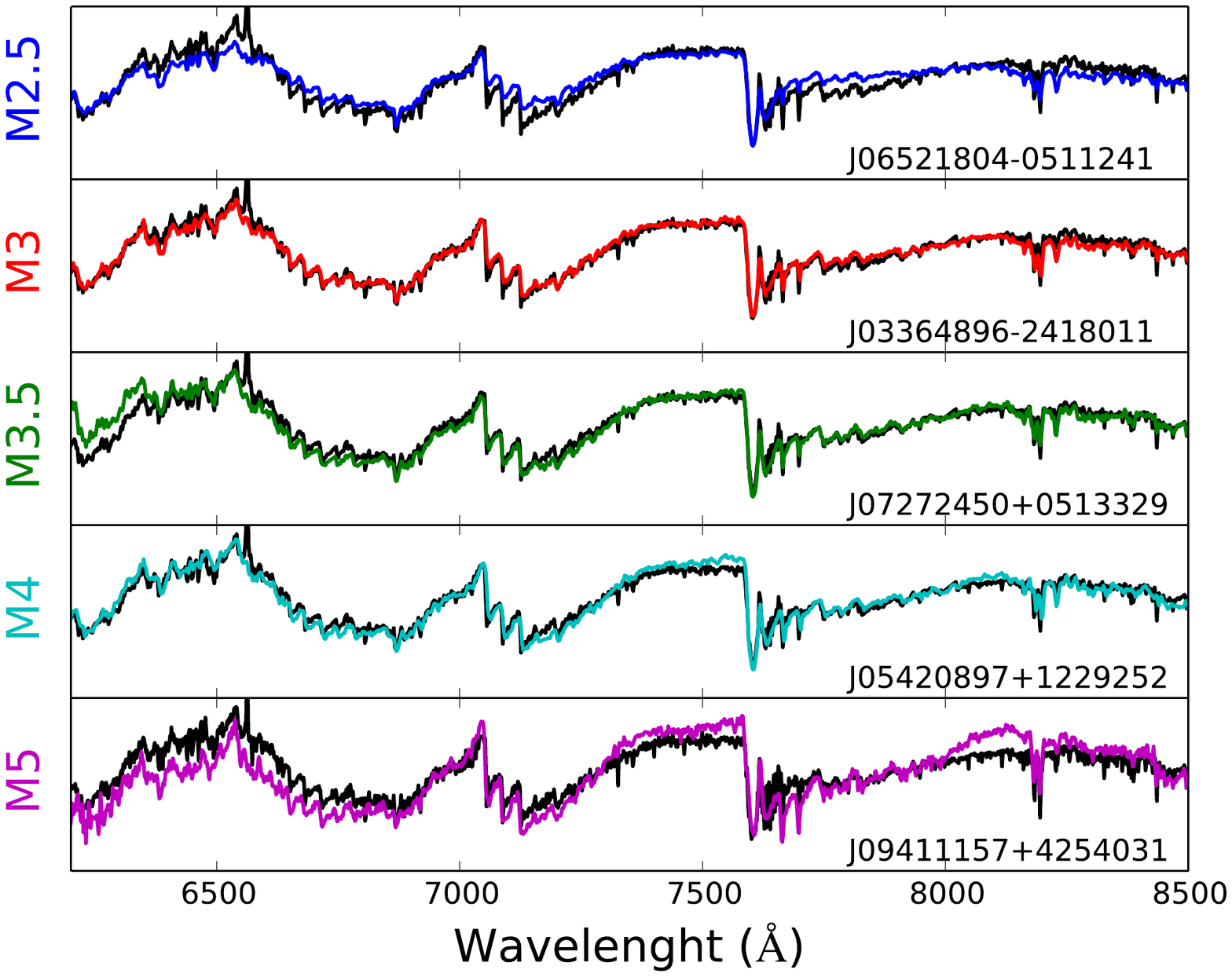}}\label{TS_1219A}
    \subfigure[\objashort~A]{\includegraphics[width=0.488\textwidth, clip=true, trim= 0cm 5cm 0cm 5cm]{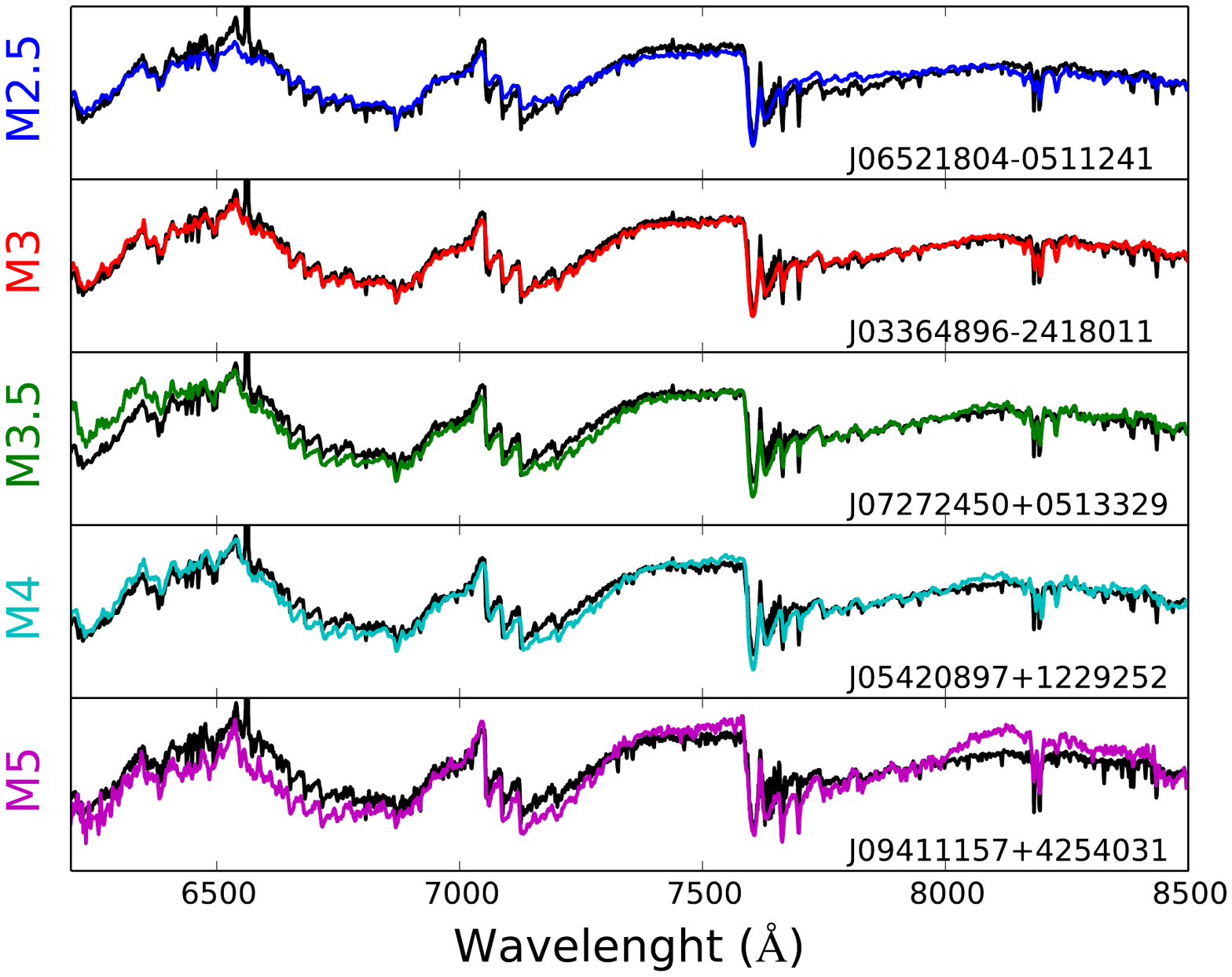}}\label{TS_2250A}
    \caption{\bf{Optical spectra} of \objbshort~A and \objashort~A (black lines) compared with known M2.5--M5 dwarfs spectra (colored lines). \bf{Colored lines corresponds to the spectral types specified on the left side of the figure panels.} The ESPaDOnS spectra were convolved and re-sampled to match the spectral resolution and dispersion relation of the literature spectra, and all spectra were normalized to a unit median. Both objects are best matched by the M3 dwarf 2MASS~J03364896--2418011.}
    \label{TS_optical}
\end{figure*}

The spectral types of the two primary stars were inferred from the TiO5 index defined by \cite{1995AJ....110.1838R}. We obtain TiO5 = 0.494 (M2.9 $\pm$ 0.5) for \objbshort\ and 0.467 (M3.2 $\pm$0.5) for \objashort. The optical spectra of \objbshort~A and \objashort~A are compared with those of known M2.5--M5 stars \citep{2003AJ....126.2421C,2007AJ....133..439C} in Figure~\ref{TS_optical}. Both spectra are visually best matched by the M3 star 2MASS~J03364896--2418011, consistently with the TiO5 indices. We therefore adopt a spectral type of M3$ \pm $0.5 for both primaries.

The radial velocities of both primaries were measured by cross-correlating their ESPaDOnS spectra with artificially Doppler-shifted slow-rotator radial velocity standards, as described by \citeauthor{2014ApJ...788...81M} (\citeyear{2014ApJ...788...81M}; see Section~4.1). This yielded heliocentric radial velocities of $5 \pm 0.5$\,\kms\ and $-16 \pm 1$\,\kms\ for \objbshort~A and for \objashort~A, respectively.

Radial velocities were also measured individually for \objbshort~A and \objbshort~B from their FIRE $H$-band spectra with a method similar to that of \cite{2017ApJS..228...18G,2015ApJ...808L..20G}. The observed spectra were compared with zero-velocity CIFIST~2011 BT-Settl spectra \citep{2012RSPTA.370.2765A,2015AA...577A..42B} with a custom IDL routine based on the amoeba Nelder-Mead downhill simplex algorithm \citep{Nelder:1965in}. A forward model was built by convolving the BT-Settl models with a one-parameter Gaussian instrumental profile, applying a Doppler shift and a two-parameters multiplicative linear slope correction to account for instrumental systematics. No telluric model was used as the observed spectra were already corrected for telluric absorption, and the Firehose pipeline already corrected the spectra for barycentric velocity variations and generates a vacuum wavelength solution. The Nelder-Mead method was used to locate the four best-fitting parameters that minimize the reduced $\chi^2$ in fifteen 0.02\,$\mu$m sections regularly distributed in the range 1.5100--1.5535\,$\mu$m to account for systematics and bad pixels, using models in the range $\log g$ = 3.0--5.5 and \teff\ = 1000--4000\,K. The median and standard deviation of the fifteen parameter measurements were adopted as the measurement and error of each individual parameter.

This method yielded heliocentric radial velocities of $-20.1 \pm 3.9$\,\kms\ for \objbshort~A and $-17.3 \pm 3.5$\,\kms\ for \objbshort~B. The FIRE radial velocity of \objbshort~A is significantly different from the $5 \pm 0.5$\,\kms\ ESPaDOnS measurement. These differing RV values could indicate that \objbshort~A is a tight binary star. No duplicated spectral lines could however be resolved in the ESPaDOnS spectrum, excluding the double-line binary (SB2) hypothesis and suggesting that it is a likely single-line binary (SB1) with a secondary significantly fainter than \objbshort~A. The equal-luminosity unresolved binary case would be detect through the spectroscopic measurement since the RV amplitude is much higher ($>25 \kms$) than the $v \sin i$ (<2\kms) measurement. 

\begin{figure*}
    \centering
    \subfigure[\objbshort~A]{\includegraphics[width=0.488\textwidth, clip=true, trim= 0cm 5cm 0cm 5cm]{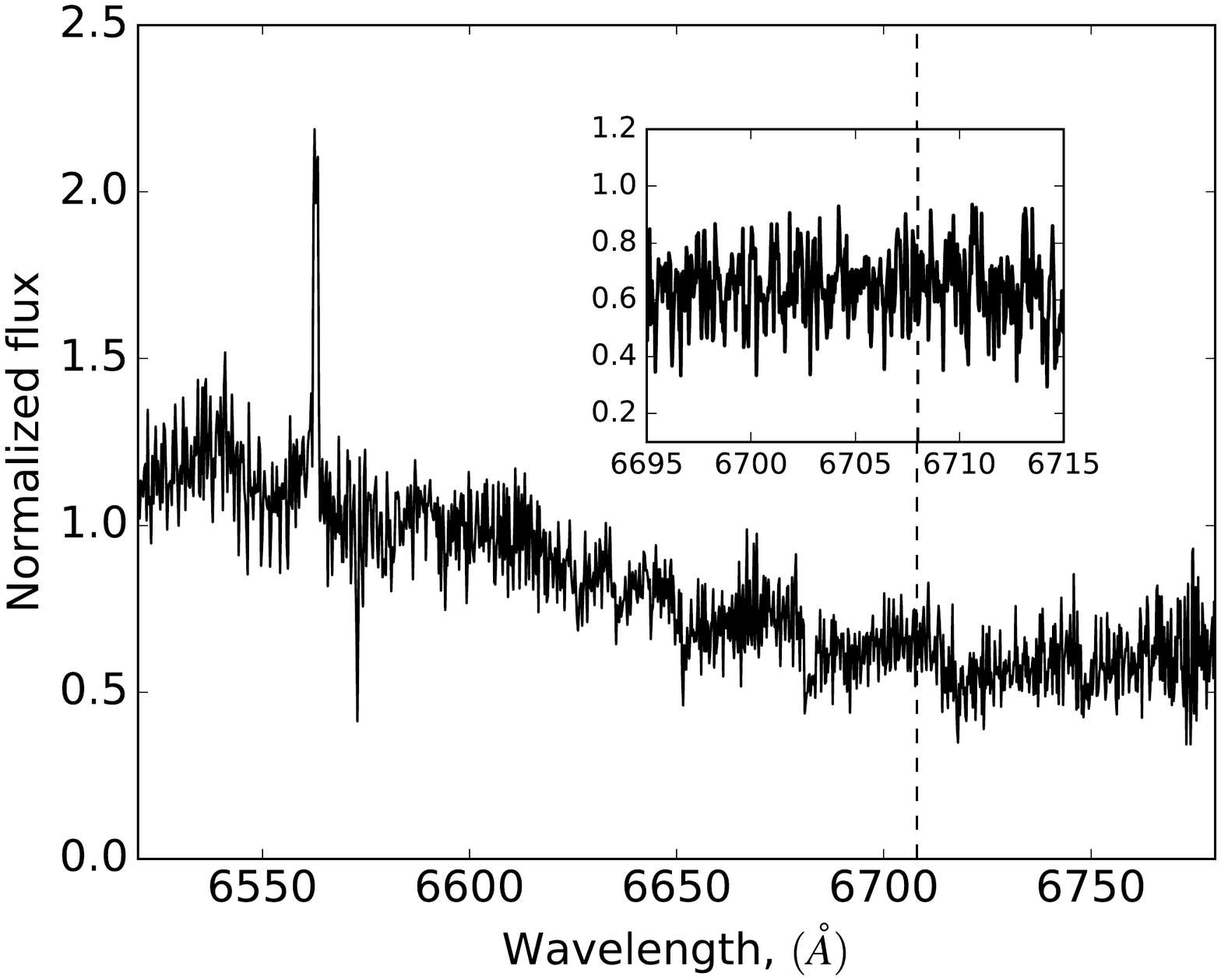}\label{1219Aspec}}
 	\subfigure[\objashort~A]{\includegraphics[width=0.488\textwidth, clip=true, trim= 0cm 5cm 0cm 5cm]{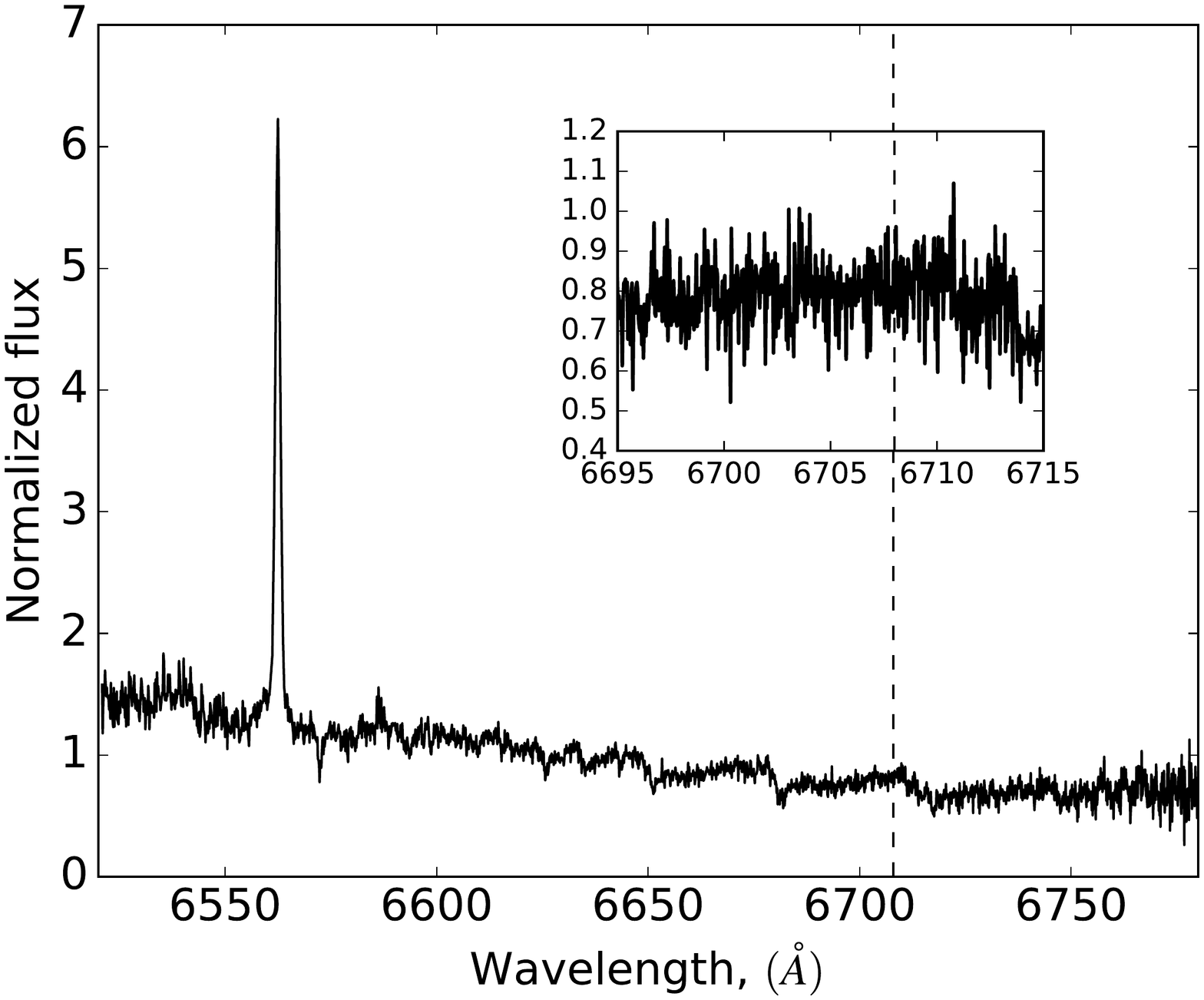}\label{2250Aspec}}
	\caption{Full-resolution ESPaDOnS spectral flux density of \objbshort~A and \objashort~A in arbitrary units. The H$\alpha$ emission lines at 6562.8\,\AA\ are clearly visible in both objects, but no lithium absorption is detected at 6708\,\AA.}
    \label{opt_spec}
\end{figure*}

The ESPaDOnS spectra were compared to a slow-rotating standard convolved by a rotational profile using the procedure described by \citealt{2014ApJ...788...81M} to measure the projected radial velocity of both primary stars. This yielded $v \sin i \sim 30$\,\kms\ and $v \sin i < 2$\,\kms\ for \objashort~A and \objbshort~A, respectively. Both ESPaDOnS spectra are displayed in Figures~\ref{1219Aspec} and \ref{2250Aspec}, where in both cases H$\alpha$ emission is prominent, but no lithium is detected at 6707.8\,\AA\, with a 3$\sigma$ upper limit of EW$_{Li} \leq 32$\,m\AA\, for \objashort~A and $\leq 54$\,m\AA\, for \objbshort~A.

The rotation rate of a fully convective star ($< 0.4$\,$M_\odot$) gradually increases as the star ages, until it reaches a maximum at $\sim$\,100\,Myr and then slows down again \citep{2000ApJ...534..335S}. The slow projected rotation of \objbshort~A is therefore either indicative of a very low inclination or an old age. 

The lack of 6707.8\,\AA\ lithium absorption in the ESPaDOnS spectra of \objashort~A and \objbshort~A is indicative of an age older than $\sim$\,20\,Myr for their M3 spectral type \citep{2001RvMP...73..719B}.

Observations by \cite{2008ApJ...689.1127M} show that members of the AB Doradus moving group (ABDMG; $150_{-20}^{+50}$\,Myr; \citealt{2004ApJ...613L..65Z,2015MNRAS.454..593B}) later than $\sim$\,K7 stop displaying lithium that would be detectable at our precision. The younger Tucana-Horologium association \citep{2000AJ....120.1410T,2001ApJ...559..388Z} and $\beta$~Pictoris moving group \citep{2001ApJ...562L..87Z}, with respective ages of $45 \pm 4$\,Myr and $24 \pm 3$\,Myr \citep{2015MNRAS.454..593B}, stop displaying lithium at spectral types $\sim$\,M2 and later, which is consistent with the 20\,Myr lower age limit estimated above. The relatively fast rotation rate ($v \sin i$ $\sim 30$\,\kms) of \objashort~A however suggests that it may be young ($\sim 100$\,Myr; e.g. see Figure~5 of \citealp{2014ApJ...788...81M}).

The spectrophotometric distances of \objashort~A and \objbshort~A were estimated from their $J$-band magnitude and the spectral type--absolute magnitude relations of \citeauthor{2013MNRAS.434.1005Z} (\citeyear{2013MNRAS.434.1005Z}; see their Table~3). We obtain an absolute $J$-band magnitude of $7.54 \pm 0.52$\,mag for both and a spectrophotometric distance of $43.0 \pm 10.3$\,pc and $39.9 \pm 9.6$\,pc respectively. The uncertainty on the spectral type (0.5 subtypes) dominates the errors on both of these measurement.

\subsection{Companion Spectral Properties}
\label{companionspectre}

\begin{figure*}
    \centering
    \includegraphics[scale = 0.6]{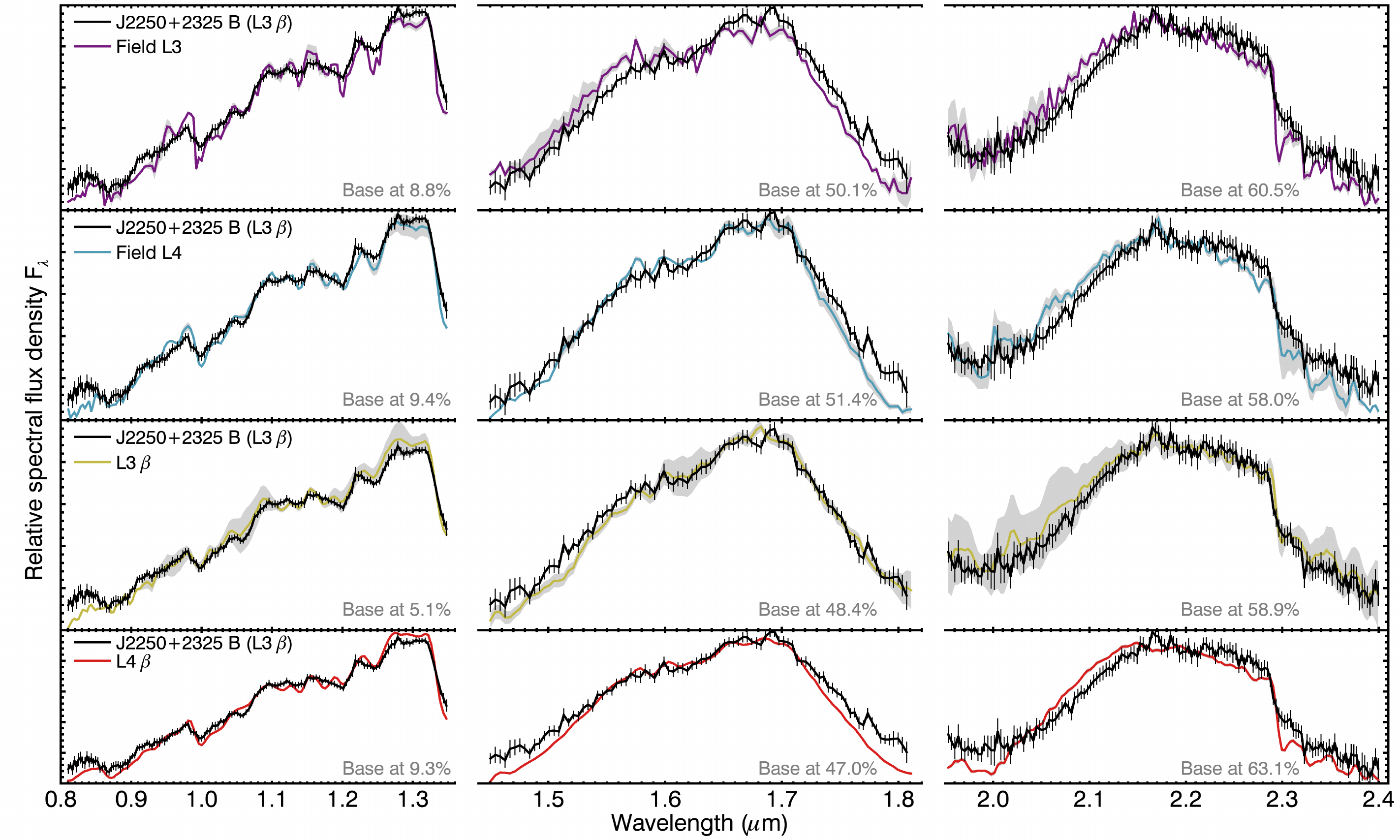}
    \caption{Comparison of the near-infrared SpeX/IRTF spectral flux density of \objashort~B (black line) compared to field and low-gravity L3 and L4 dwarf templates from \citeauthor{2015ApJS..219...33G} (\citeyear{2015ApJS..219...33G}; colored lines). The spectrum of \objashort~B was convolved at the resolution of the template spectra, and each band was normalized individually to its median value to facilitate a comparison of the spectral features. The grey regions display the scatter in individual spectra used to construct the comparison templates, and the vertical black lines indicate the error bars on the convolved \objashort~B spectrum. The grey text in each figure cell indicates how far the zero flux level lies below the window, in fractional units of the full figure cell vertical range. \objashort~B is best matched visually by the L3\,$\beta$ template.}
    \label{spectrecomp}
\end{figure*}

The spectral templates of \cite{2015ApJS..219...33G} were used to visually determine the spectral type and gravity class \citep{2005ARA&A..43..195K} of \objashort~B, as shown in Figure~\ref{spectrecomp}. The templates are built from a median combination of several objects within a same spectral subtype and gravity classed, done independently in each band ($z$, $J$, $H$ and $K$; see also K.~L.~Cruz et al., submitted to AJ\footnote{Soft release available at \url{https://github.com/kelle/NIRTemplates\_Manuscript/releases/tag/v1}}). The best-matching template was found to be L3\,$\beta$, indicating subtle signs of low gravity, which is typically consistent with an age of $\sim$\,100\,Myr \citep{2009AJ....137.3345C}. This age estimate is consistent with that of \objashort~A based on its projected rotational velocity. The index-based classification scheme of \cite{2013ApJ...772...79A} yields a spectral type of L2.5$ \pm 0.6$ with a gravity class INT-G and gravity score `1011' (see \citealp{2013ApJ...772...79A} for more detail) , which is consistent with L3\,$\beta$.

The FIRE spectrum of \objbshort~B was also visually compared with the spectral templates of \cite{2015ApJS..219...33G}, and was found to be a best match with the field M9 spectral type, with no visible peculiarity or signs of low gravity. The index-based spectral classification of \cite{2013ApJ...772...79A} yields a consistent M9 FLD-G classification, with a gravity score `0n10'. Such a lack of low-gravity spectral signatures indicates that the \objbshort~AB system is likely older than $\sim$\,200\,Myr \cite{2013ApJ...772...79A}, which is consistent with the low projected rotational velocity of \objbshort~A.

The spectrophotometric distance of \objashort~B was estimated with the spectral type--absolute magnitude relation of \citeauthor{2015ApJ...810..158F} (\citeyear{2015ApJ...810..158F}; see their Table~10). This yielded an absolute $J$-band magnitude of $12.874 \pm 0.402$\,mag, which was compared with our MKO $J$-band measurement converted to the 2MASS system ($16.845 \pm 0.02$\,mag) with the relations of \citealt{2006MNRAS.373..781L}, resulting in a distance estimate of $62.3 \pm 11.4$\,pc. The estimated spectrophotometric distance of \objbshort~B is $52.2 \pm 9.7$\,pc using the same procedure. The uncertainty on spectral type (0.5 subtypes) dominates the measurement error.

\subsection{Common Proper Motion}\label{seccpm}

This section describes a verification that both new candidate companions discovered in this survey are co-moving with their respective host star. 

\begin{figure}
    \centering
    \includegraphics[scale=0.5]{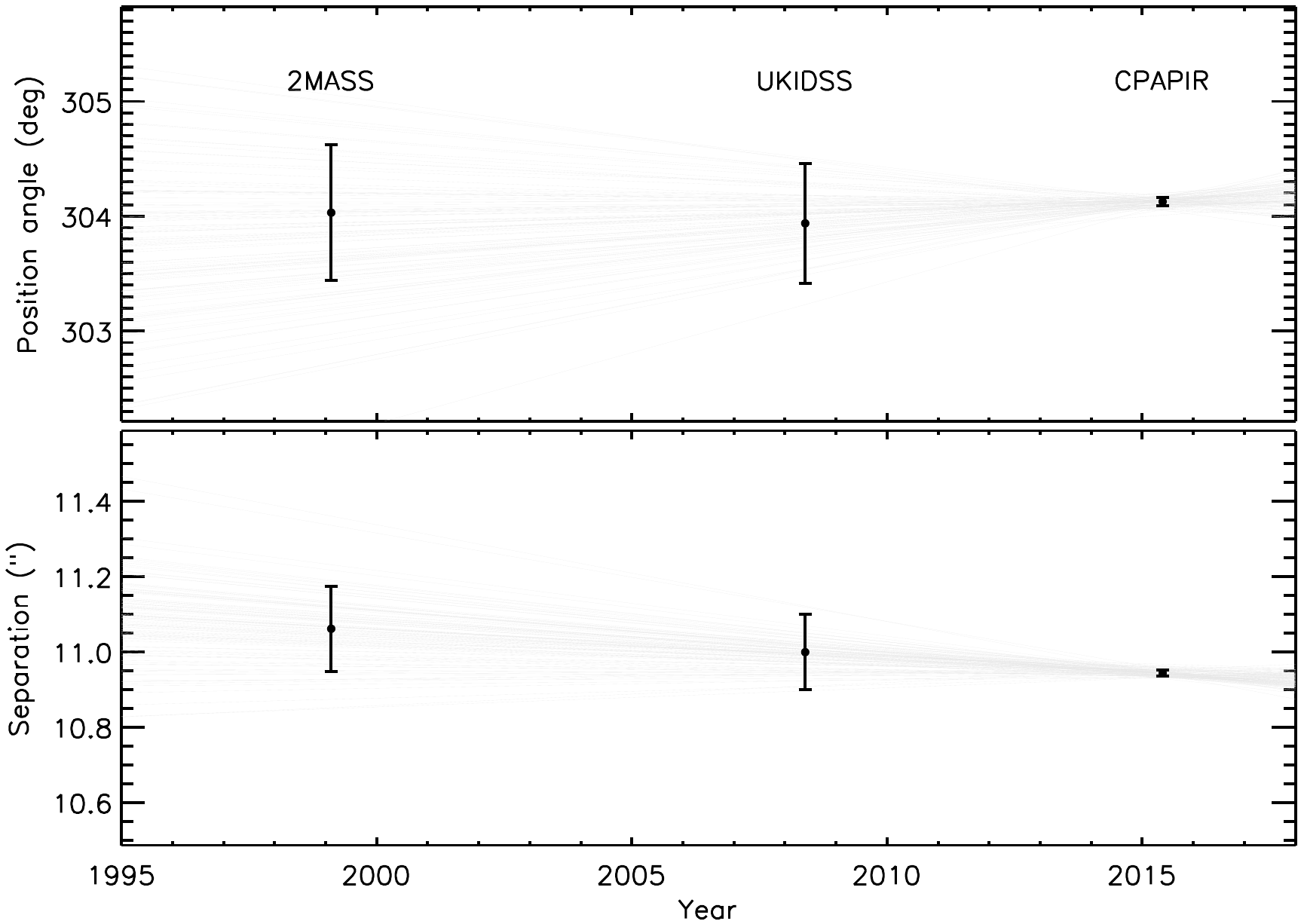}
    \caption{Separation and position angle of \objbshort~A and \objbshort~B at the three available epochs. A hundred random draws of the posterior probability distribution in separation and position angle temporal evolution are displayed as gray lines. These measurements are consistent with \objbshort~A and B being co-moving.}
    \label{sep1219}
\end{figure}

The proper motion of \objbshort~A was reported by \cite{2012AJ....143...80S} and is listed in Table~\ref{alll}. The proper motion of \objbshort~B was measured from its detections in UKIDSS \citep{2007MNRAS.379.1599L}, 2MASS $J$-band, and 2015 CPAPIR $J$-band, using a 100-elements Monte Carlo method where a linear slope is adjusted through three synthetic astrometric measurements drawn from the posterior probability distribution of the astrometry at each epoch. The position angle of the companion is $304.13 \pm 0.04$\,\textdegree\ with a variation of $0.01 \pm 0.01$\,\textdegree\,$yr^{-1}$, and its separation is $10.943 \pm 0.008$\arcsec\ with a variation of $-7.3 \pm 3.0$\,\masyr, as shown in Figure~\ref{sep1219}. If \objbshort~B was a fixed background star, its separation from the primary star would have increased by $1.185 \pm 0.060$\arcsec\ across the 12\,yr separating the 2MASS and CPAPIR astrometric measurements, and would have a position angle of $310.29 \pm 0.35$\,\textdegree\ at the CPAPIR epoch, which is inconsistent with our measurement.

The common proper motion and the small separation of \objbshort~A and \objbshort~B strongly suggest that the system is gravitationally bound. The fact that the FIRE radial velocities of \objbshort~A and \objbshort~B are similar is expected if they are co-moving, however it is not necessarily expected given that we identified \objbshort~A as a potential radial velocity variable itself from the different FIRE and ESPaDOnS radial velocity measurements with 30 days delay. Determining whether the discrepant measurements are caused by instrumental systematics or by the primary star being a binary will require additional measurements. However, if \objbshort~A is an unresolved binary with the secondary 6 times ($\sim 2$ mag) fainter than the primary, our spectrophotometric distance estimate becomes $43 \pm 10\,pc$, which is consistent with that obtained for \objbshort~B ($52.2 \pm 9.7$).

\begin{figure}
    \centering
    \includegraphics[scale = 0.5]{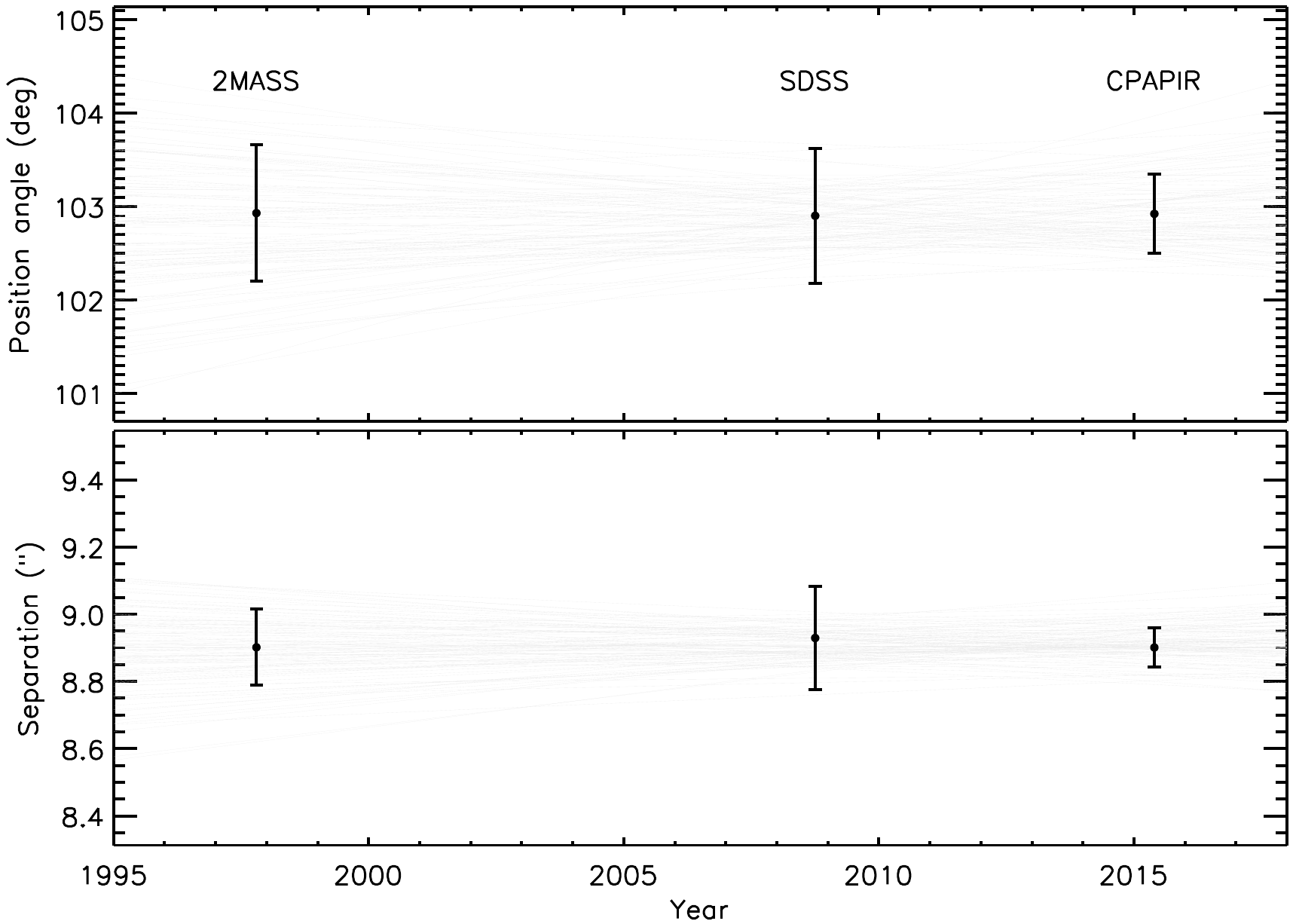}
    \caption{Separation and position angle of \objashort~A and \objashort~B at the three available epochs. A hundred random draws of the posterior probability distribution in separation and position angle temporal evolution are displayed as gray lines. These measurements are consistent with \objbshort~A and B being co-moving.}
    \label{pa_sep}
\end{figure}

\cite{2015AJ....150..137Q} report a proper motion for \objashort~A, which is listed in Table~\ref{alll}. A similar method than that described above was used to measure the motion of \objashort~B relative to the primary, using 2MASS $J$, 2005 and 2009 SDSS $z$, and 2015 CPAPIR $J$-band detections. Both the separation and position angle were found to remain constant at $8.90 \pm 0.06$\arcsec\ and $102.9 \pm 0.4$\textdegree, with upper limits of $5$\,\masyr\ and $0.033 $\textdegree\,yr$^{-1}$ on their variation as shown in Figure~\ref{pa_sep}. If \objashort~B was a fixed background star, its separation would have varied by $1.0843 \pm 0.0015$\arcsec\ during the 12\,yr separating its 2MASS and CPAPIR detections. The expected position angle at the CPAPIR epoch would also be at $279.3 \pm 0.5$\,\textdegree.

Although we do not have a radial velocity measurement for the companion, the close proximity and the common proper motion of \objashort~A and \objashort~B suggest that they are gravitationally bound.

Using the 265 nearby ultracool dwarfs proper motion measurements of \cite{2012ApJS..201...19D}, we infer a median tangential velocity of 31.8\,\kms\ for their population. Assuming a normal isotropic distribution of substellar-mass objects, we find that there is only a $\sim$\,1\% probability that a L3$\beta$ dwarf randomly falls this close to a random field star with a common proper motion consistent at $< 2\sigma$ with our measurements.

\begin{figure}
    \centering
    \includegraphics[clip=true, trim= 1cm 3cm 0cm 3cm,  scale=0.45]{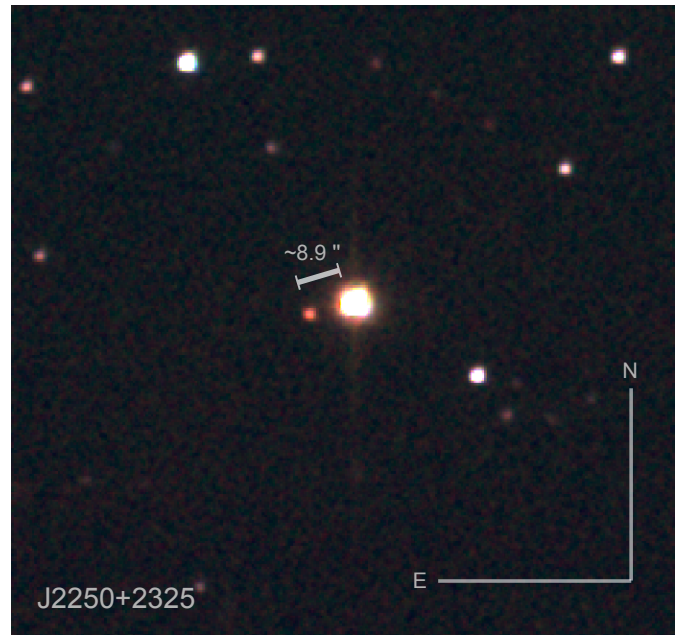}
    \caption{Color image of the \objashort~AB system built from the CPAPIR $J$ (blue), $H$ (green) and $K$ (red) images. \objashort~B appears much redder than background stars in the field of view.}
    \label{image2250}
\end{figure}

Figure~\ref{image2250} shows the combined $J$, $H$ and $K$-band of 2016 CPAPIR images of the \objashort~AB system, and demonstrates that the companion is significantly redder than other stars in the field. The angular separation and position angle of the companion appears very similar to the 2MASS images (see Figure~\ref{psf2250}) obtained 13\,yr earlier.

\section{MEMBERSHIP TO YOUNG ASSOCIATIONS}\label{sec:member}

\objashort~A was included in our sample of suspected young stars because it was listed as a candidate member of ABDMG by \cite{2012AJ....143...80S}, in a survey based on proper motion, X-ray and UV activity indicators.

An age of $\sim$\,50\,Myr was initially estimated for ABDMG \citep{2004ARAA..42..685Z} based on the H$\alpha$ emission and the color-magnitude position of its M-type members. \cite{2013ApJ...766....6B} subsequently found an inconsistent lower limit of 110\,Myr based on the pre-main sequence contraction times for K-type members and the fact that they appeared to have reached the main sequence in ABDMG. Later studies such as \cite{2014ApJ...792...37M} demonstrated that the strong magnetic activity of early M dwarfs made them appear younger in a color-magnitude diagram when compared to non-magnetic models, which explains why early estimates of moving group ages were too low. More recently, \cite{2015MNRAS.454..593B} derived an isochronal age of $150_{-20}^{+50}$\,Myr for ABDMG.

\begin{figure*}
    \centering
    \includegraphics[scale=0.28]{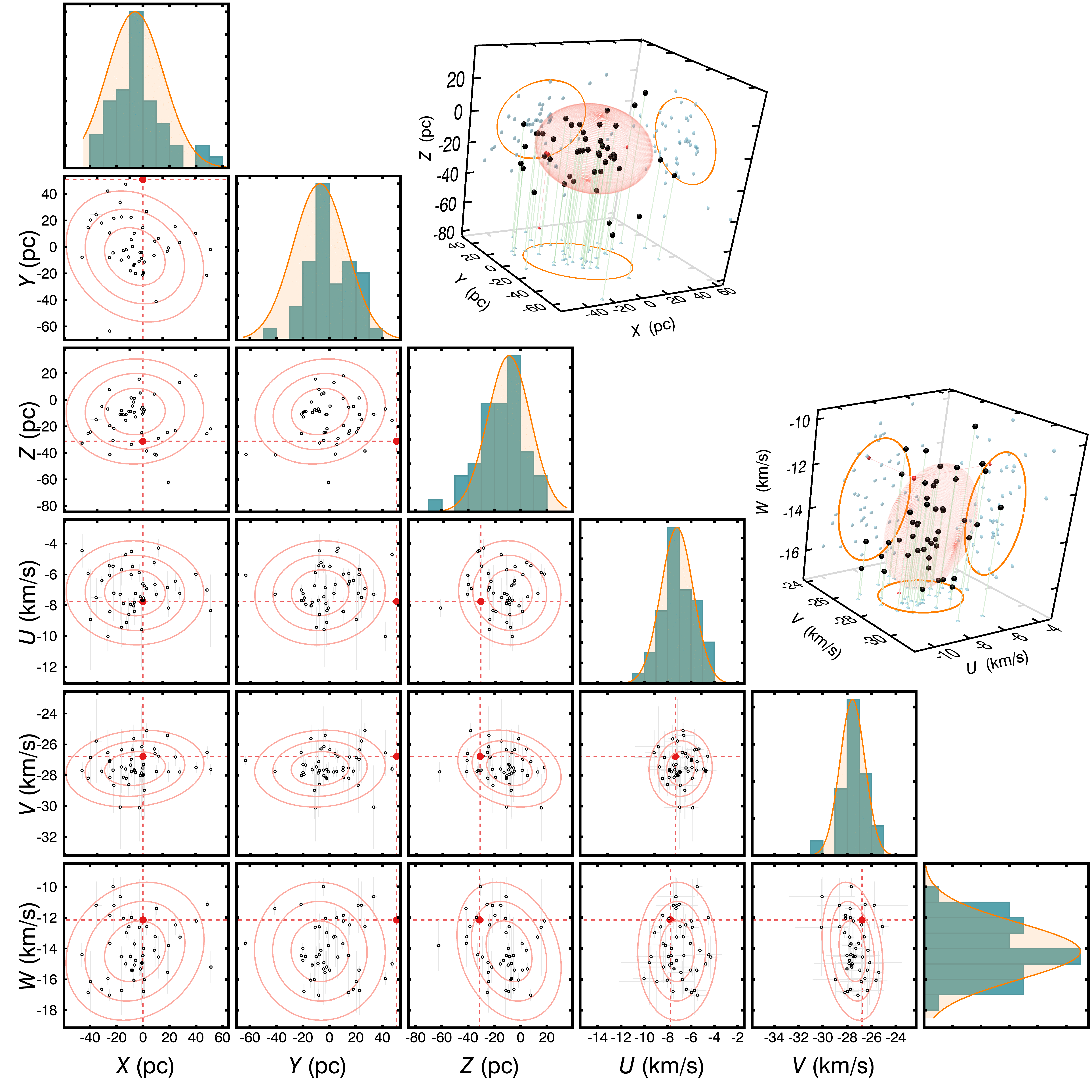}
    \caption{Predicted $XYZ$ Galactic position and $UVW$ space velocity of \objashort~A (red circles) obtained by adopting the statistical kinematic distance that maximizes the BANYAN~II membership probability. Known members of ABDMG are represented with black circles, and orange thick lines show the multivariate Gaussian model of ABDMG used in BANYAN~$\Sigma$, which is an updated version of the BANYAN~II model (Gagn\'e et al., submitted to ApJS). The 2D projections show 1, 2 and 3$\sigma$ contours of the model and the 1D and 3D projections show a 1$\sigma$ contour only. 1D projections of the ABDMG models are also compared with histograms of the member positions in $XYZUVW$ space.}
    \label{ellipsoid}
\end{figure*}

We used the BANYAN~II tool \citep{2014ApJ...783..121G,2013ApJ...762...88M} to assess the possibility that \objashort~A is a member of a nearby young moving group. We used the `young' mode of BANYAN~II where the star is assumed to be younger than 1\,Gyr, and included the proper motion and radial velocity measurements listed in Table~\ref{alll}. We obtained a 5.57\% Bayesian probability that the system is a member of ABDMG, with a statistical distance of $58.2 \pm 4.2$\,pc (assuming membership). Figure~\ref{ellipsoid} shows the projected position of \objashort~A in $XYZ$ Galactic position and $UVW$ space velocities compared to other members of ABDMG and the BANYAN~II Gaussian model of ABDMG.

The low Bayesian probability for the ABDMG membership can be assigned to the fact that the $Y$ position of \objashort~A is more than 3$\sigma$ away from the locus of ABDMG members. 

\citealt{2017AJ....153...18B} recently identified 2MASS~J22362452+4751425~AB, a young system similar to \objashort~AB that lies in a similar $Y$ region away from ABDMG, and they hypothesize that the true spatial distribution of ABDMG members may be larger than currently known. The discovery of most known ABDMG members was based on the Hipparcos survey \citep{1997A&A...323L..49P}, which would not have been able to identify members at the position where \objashort~AB and 2MASS~J22362452+4751425~AB are located.

We investigated the probability that a random young system such as \objashort~AB falls this close to ABDMG members in $XZ$ and $UVW$ while ignoring the $Y$ dimension, without being related to the ABDMG moving group. $10^7$ synthetic objects were drawn from the distribution of young stars ($<$\,250\,Myr) within 58\,pc of the \besancon\ Galactic Model \citep{1996AA...305..125R,2003AA...409..523R,2012AA...538A.106R,2016AN....337..884R}. The \besancon\ model indicates that the fraction of stars younger than 250\,Myr in the Solar neighborhood is of $\sim$\,1.98\%, from which we adjusted the spatial density of low-mass stars measured by \cite{2003PASP..115..763C}, yielding a density of $9.5 \times 10^{-3}$\,$pc^{-3}$. The $N\sigma$ distance from the BANYAN~II ABDMG model was calculated for all synthetic objects (ignoring $Y$), and for \objashort~A. We find that a total of $0.97 \pm 0.04$ young M dwarfs are expected to be found within the $N\sigma$ distance (1.4) of \objashort~A. This indicates that there is a 62\% probability that \objashort~A is a random interloper to the ABDMG distribution.

Including the fact that 2MASS~J22362452+4751425~AB is also detected near the ABDMG distribution in this simulation, we obtain a probability of 38\% that both objects are kinematic interlopers unrelated to ABDMG. It is therefore not possible at this stage to determine with confidence whether these systems are part of a yet unknown spatial extension of ABDMG. Trigonometric distance measurements of both primary stars will help strengthen the case for their ABDMG membership, as well as additional searches for new members in this region of the Solar neighborhood.

Using BANYAN from \cite{2013ApJ...762...88M} with consideration of the 2MASS $J$-band and APASS $i'$-band (12.815 mag) photometric data \citep{2016yCat.2336....0H} and radial velocity, the membership's probability to ABDMG goes up to 99.91\% ; 19.60\% as a single star and 80.31\% as a binary star at a statistical distance of $60 \pm 3.8$\,pc. \objashort~A is $\sim 0.75$ mag too bright on the $M_J$ vs $I_c - J$ color-magnitude diagram assuming a distance of 60\,pc to match the single ABDMG members track (see Figure 3 of \citealt{2013ApJ...762...88M}) and the presence of an SB1 could explained this excess. The spectrophotometric distance deduced in Section~\ref{sec:Results} is also an indication of a binary star. Assuming that \objashort~A is actually 2 stars of equal brightness, the spectrophotometric distance should be increase by $\sqrt 2$, resulting in $60.8 \pm 14.6$\,pc. This distance is exactly what it is expected for the system considering its membership to ABDMG. This distance is also more consistent with that deduced for \objashort~B ($62.3 \pm 11.4$\,pc). A radial velocity or adaptive-optic follow-up on \objashort~A should provide a strong constraint on the nature of that star. These membership probabilities are high in comparison with them from BANYAN~II because groups are modeled with different 3D Gaussian ellipsoids in XYZ coordinates.

\objbshort~A was similarly reported as a candidate member of ABDMG by \cite{2012AJ....143...80S}. Using BANYAN~II tool with the proper motion listed in Table~\ref{alll}, we find a Bayesian probability of 0.1\% for ABDMG membership, and 81.9\% for TW~Hya membership ($10 \pm 3$\,Myr; \citealp{1989ApJ...343L..61D,1997Sci...277...67K,2015MNRAS.454..593B}) when the `young' mode is used. If no age is assumed for the system, the respective ABDMG and TW~Hya probabilities fall to 0\% and 0.01\%. The low projected rotational velocity and absence of lithium in the ESPaDOnS spectrum of \objbshort~A, and the lack of low-gravity features in the SpeX spectrum of \objbshort~B provide strong indications that it is most likely not a member of a young association. Whether the radial velocities from ESPaDOnS and FIRE are included or not, both BANYAN~I and II yield a 0\% membership probability to any young moving group for \objbshort.

\begin{deluxetable*}{lccccl}[p]
\tablecaption{Properties of the two companions discovered in this survey. \label{alll}}
\tablehead{\colhead{2MASS Designation} & \multicolumn{2}{c}{12193316+0154268} & \multicolumn{2}{c}{22501512+2325342} & \colhead{Ref.}}
\startdata
Component & A & B & A & B & \\
R.A. (hh:mm:ss) & 12:19:33.157 & 12:19:32.541 & 22:50:15.12 & 22:50:15.766 & 1\\
Decl. (dd:mm:ss) & 01:54:26.77 & 01:54:32.69 & 23:25:34.2 & 23:25:32.64 & 1\\
\hline
Association & \multicolumn{2}{c}{Field} & \multicolumn{2}{c}{ABDMG} & 1\\
Age & \multicolumn{2}{c}{$>$\,1\,Gyr} & \multicolumn{2}{c}{$\sim$\,150\,Myr} & 1\\
Kin. distance\tablenotemark{a} (pc) & \multicolumn{2}{c}{$\cdots$} & \multicolumn{2}{c}{$58.2 \pm 4.2$} & 1\\
Spec. distance (pc) & $39.9 \pm 9.6$ & $52.2 \pm 9.7$ & $43.0 \pm 10.3$ & $62.3 \pm 11.4$ & 1\\
Spec. distance if binary (pc) & $43 \pm 10$ & $\cdots$ & $60.8 \pm 14.6$ & $\cdots$ & 1\\ 
Angular separation (\arcsec) & \multicolumn{2}{c}{$10.943 \pm 0.008$} & \multicolumn{2}{c}{$8.90 \pm 0.06$} & 1\\
Proj. Phys. separation (AU) & \multicolumn{2}{c}{$571 \pm 106$\tablenotemark{b}} & \multicolumn{2}{c}{$518 \pm 40$} & 1\\
Position angle (\textdegree) & \multicolumn{2}{c}{$304.13 \pm 0.04$} & \multicolumn{2}{c}{$102.9 \pm 0.4$} & 1\\
$\mu_\alpha\cos\delta$ (\masyr) & \multicolumn{2}{c}{$-73.40 \pm 4.7$} & \multicolumn{2}{c}{$68.7 \pm 1.2$} & 1,2,3\\
$\mu_\delta$ (\masyr) & \multicolumn{2}{c}{$-66.10 \pm 5.4$} & \multicolumn{2}{c}{$-58.7 \pm 1.9$} & 1,2,3\\
Heliocentric Radial velocity (\kms) & & &  &  & \\
\hspace{5mm}2015 May 31 (UTC) & $-20.1 \pm 3.9$ & $-17.3 \pm 3.5$ & $\cdots$ & $\cdots$ & 1\\
\hspace{5mm}2015 Jul 1 (UTC) & $\cdots$ & $\cdots$ & $-16 \pm 1$ & $\cdots$ & 1\\
\hspace{5mm}2015 Jul 5 (UTC) & $5 \pm 0.5$ & $\cdots$ & $\cdots$ & $\cdots$ & 1\\
$v \sin i$ (\kms) & $<2 $& $\cdots$& $ \sim 30$ &$\cdots$& 1\\
Spectral type & M3 & M9 & M3 & L3\,$\beta$ & 1\\\hline
SDSS $u^\prime$ & $18.25 \pm 0.02$ & $22.4 \pm 0.5$ & $17.88 \pm 0.02$ & $24 \pm 3$ & 4\\
SDSS $g^\prime$ & $15.465 \pm 0.003$ & $23.6 \pm 0.4$ & $15.599 \pm 0.003$ & $23.9 \pm 0.7$ & 4\\
SDSS $r^\prime$ & $14.074 \pm 0.004$ & $21.8 \pm 0.1$ & $14.150 \pm 0.002$ & $23.3 \pm 0.7$ & 4\\
SDSS $i^\prime$ & $15.44 \pm 0.01$ & $19.10 \pm 0.02$ & $13.069 \pm 0.005$ & $24 \pm 2$ & 4\\
SDSS $z^\prime$ & $11.962 \pm 0.003$ & $17.36 \pm 0.02$ & $12.159 \pm 0.003$ & $19.5 \pm 0.1$ & 4\\
DENIS $I$ & $11.97 \pm 0.03$ & $18.3 \pm 0.2$ & $\cdots$ & $\cdots$ & 5\\
2MASS $J$ & $10.54 \pm 0.02$ & $\cdots$ & $10.71 \pm 0.02$ & $\cdots$ & 6\\
2MASS $H$ & $9.96 \pm 0.02$ & $\cdots$ & $10.08 \pm 0.02$ & $\cdots$ & 6\\
2MASS $K_S$ & $9.66 \pm 0.01$ & $\cdots$ & $9.83 \pm 0.02$ & $\cdots$ & 6\\
CPAPIR MKO $J$ & $\cdots$ & $\cdots$ & $\cdots$ & $16.71 \pm 0.03$ & 1\\
CPAPIR MKO $H$ & $\cdots$ & $\cdots$ & $\cdots$ & $15.71 \pm 0.05$ & 1\\
CPAPIR MKO $K$ & $\cdots$ & $\cdots$ & $\cdots$ & $14.89 \pm 0.04$ & 1\\
UKIDSS $J$ & $10.652 \pm 0.001$ & $14.977 \pm 0.004$ & $\cdots$ & $\cdots$ & 7\\
UKIDSS $H$ & $11.151 \pm 0.001$ & $14.312 \pm 0.005$ & $\cdots$ & $\cdots$ & 7\\
UKIDSS $K$ & $10.716 \pm 0.001$ & $13.787 \pm 0.005$ & $\cdots$ & $\cdots$ & 7\\
AllWISE $W1$ & $9.54 \pm 0.02$ & $\cdots$ & $9.72 \pm 0.02$ & $\cdots$ & 8\\
AllWISE $W2$ & $9.40 \pm 0.02$ & $\cdots$ & $9.58 \pm 0.02$ & $\cdots$ & 8\\
AllWISE $W3$ & $9.27 \pm 0.04$ & $\cdots$ & $9.44 \pm 0.04$ & $\cdots$ & 8\\
\enddata
\tablerefs{(1)~This paper, (2)~\citealt{2012AJ....143...80S}, (3)~\citealt{2015AJ....150..137Q}, (4)~\citealt{2015ApJS..219...12A}, (5)~\citealt{1997Msngr..87...27E}, (6)~\citealt{2006AJ....131.1163S}, (7)~\citealt{2007MNRAS.379.1599L}, (8)~\citealt{2014ApJ...783..122K}.}
\tablenotetext{a}{Kinematic distance based on moving group membership.}
\tablenotetext{b}{Using the companion's spectrophotometric distance.}
\end{deluxetable*}

\section{DISCUSSION}
\label{discussion}

We estimate the 1$\sigma$ lower limit of the recovery efficiency of our survey to $0.341^{1/5}$ = 80.6\% based on the fact that the 5 known companions within our selection criteria were recovered. Even if the input sample used for this survey is probably highly incomplete, we can estimate a lower limit for the occurrence of 3--18\arcsec\ substellar companions to young M dwarfs at 6/2812 $\approx$ 0.21\%. This lower limit is consistent with the results of \citeauthor{2016PASP..128j2001B} (\citeyear{2016PASP..128j2001B}; see their Table~3), where the frequency of 5--13\,\Mjup\ planetary-mass objects on wide orbits (100--1000\,AU) around M-type stars is estimated at <\,7.3\%, based on a compilation of multiple direct-imaging surveys. For comparison, the occurrence rate of all substellar-mass companions (2--80\,\Mjup) on 8--400\,AU orbits was estimated at $4.4^{+3.2}_{-1.3}$\% \citep{2016AA...596A..83L}, based on a Bayesian analysis of direct-imaging observations of 54 stars.
 
\begin{figure}
    \centering
    \includegraphics[scale=0.32, angle=90]{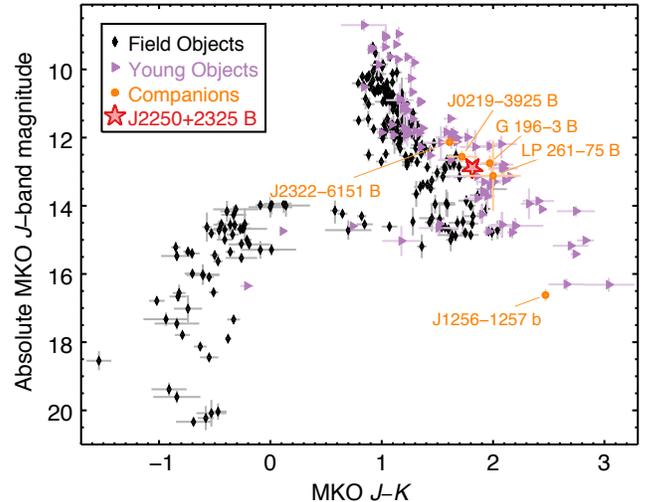}
    \caption{Near-infrared color-magnitude diagram of low-mass stars and substellar objects. Field objects are displayed as black diamonds, young objects are displayed as purple rightward triangles, and the known companions recovered in this survey are displayed as orange circles. \objashort~B is displayed as a thick red star, and falls in the locus of young L dwarfs that have redder near-infrared colors than the field sequence.}
    \label{cmd}
\end{figure}

The fsed$=2$ cloudy models of \cite{2012ApJ...750...74S} were used to estimate a mass of $30_{-4}^{+11}$\,\Mjup\ for \objashort~B, assuming the age of ABDMG and a temperature of $1600 \pm 200$\,K (see Table~19 of \citealt{2016ApJS..225...10F}). The age of the system dominates the error bars reported here, but the poorly constrained systematics of models at these young ages and low masses are likely to be at least as large.

We estimate a large mass ratio of $q \sim 0.08$ for the \objashort~AB system, similar to other systems found by direct imaging (e.g., 2MASS~12073346--3932539~Ab at $q \sim 0.15$; \citealt{2004AA...425L..29C}). Such high mass ratios are most likely associated with systems in which both components are formed through turbulent core fragmentation. Assuming a circular orbit and a face-on orientation, we estimate a total period of $\sim$\,19200\,yr for the system. The orbital motion would therefore induce an orbital motion of 3\,\masyr. The sensitivity of \textit{Gaia} will be of the order of 130\,$\mu$as for late-type, faint objects ($V$=20\,mag)\footnote{Science performance from In-Orbit Commissioning Review (July 2014) available at \url{http://www.cosmos.esa.int/web/gaia/science-performance}}, making it likely possible to measure the orbital motion of \objashort~B in the near future. This observation would help to constrain the mass ratio of the system independently from evolutionary models. With a precise mass estimation, evolutionary models could provide a better constraint on the age of the system, which will be useful to assess its possible membership to ABDMG.

We suggest that \objashort~A could be a binary star due to its brightness excess compared to single member of ABDMG. It would not be the only low-mass binary system with a companion on a relatively wide orbit (e.g.  2MASS~J01033563--5515561ABb \citep{2013A&A...553L...5D}, Ross~458ABc \citep{2010MNRAS.405.1140G}, 2MASS~J1256--1257 \citep{2016ApJ...818L..12S}) and suggests, again, that the energies exchange in a three bodies system is most likely to produce a tight binary system with the ejection of the third and less massive component resulting in a distant orbiting companion or an escape as mentioned by \cite{2015AJ....149..145R} for binary BD. We could also mention Gu~PscAb \citep{2014ApJ...787....5N} as a potential binary system hosting a planetary-mass object on a very wide orbit. Gu~Psc~A was found to be a member of ABDMG through youth signature, activity and high Bayesian analysis probability (using BANYAN). However, GU~Psc~A appeared $\sim 0.78$ mag brighter than the absolute magnitude predicted for a single ABDMG member with its color \citep{2013ApJ...762...88M} in a very similar way as \objashort~A. Since the multi-epoch radial velocity measurements remain constant, they could not conclude that GU~Psc~A is a binary system and suggest that the excess could be a consequence of chromospheric activity. However, the binary nature of GU~Psc ~A remains a possibility if the system is in a near pole-on geometry.\\

The near-infrared color-magnitude positions of the six co-moving pairs recovered in this survey is displayed in Figure~\ref{cmd}. It can be noted that \objashort~B and the other previously known young companions fall in the locus of young L-type objects which are redder than the sequence of field low-mass stars and brown dwarfs.

\section{SUMMARY AND CONCLUSIONS}
\label{conclusion}

The discovery of a red ($J - K = 1.814$) L3\,$\beta$ co-moving companion to the M3-type potential ABDMG member \objalong, and of the new field co-moving M3+M9 pair \objblong~AB, are reported. Both primary stars were suggest to be binaries and future observations should be done to confirm it. \objashort~B is separated by $8.9 \pm 0.05$\arcsec\ from its host star, corresponding to a projected physical separation of $\sim$\,500\,AU at the kinematic distance that best matches ABDMG, and has a model-dependent mass of $\sim$\,30\,\Mjup\ assuming the age of ABDMG. The system has kinematics similar to those of ABDMG members and the companion displays features consistent with a low gravity and therefore a young age, but the $Y$ position of \objalong~AB is an outlier compared to ABDMG members, similarly to the 2MASS~J22362452+4751425~AB system discovered by \cite{2017AJ....153...18B}. We determine that there is however a relatively high probability (38\%) that two young M dwarf unrelated to ABDMG fall this close to the ABDMG locus in $XZ$ and $UVW$ coordinates, therefore their discovery alone do not provide a strong indication that ABDMG is more extended than previously known. This possibility should however warrant further study.

The discovery of \objashort~B from 2MASS data illustrates that decade-old surveys can lead to useful discoveries of peculiar objects. A similar study could be led using others data sets such as UKIDSS, Spitzer archival data, or future projects like the Large Synoptic Survey Telescope (LSST) survey \citep{2008arXiv0805.2366I}. Similar studies will also become more efficient as more young association members are compiled and new moving groups are uncovered.

\acknowledgments
This work was supported in part by grants from the Fonds de Recherche Qu\'eb\'ecois--Nature et Technologie (FRQNT), 2015 and 2016 Trottier Excellence Grants for Summer Interns iREx (Institute for research on exoplanets). This paper includes data gathered with the 6.5 meter Magellan Telescopes located at Las Campanas Observatory, Chile (CNTAC program CN2013A-135). Based on observations obtained as part of the VISTA Hemisphere Survey, ESO Program, 179.A-2010 (PI: McMahon). This publication uses observations obtained at Infrared Telescope Facility, which is operated by the University of Hawaii under contract NNH14CK55B with the National Aeronautics and Space Administration through program number 2015B027 (PI Gagn\'{e}). The authors recognize and acknowledge the very significant cultural role and reverence that the summit of Mauna Kea has always had within the indigenous Hawaiian community. We are most fortunate to have the opportunity to conduct observations from this mountain. This research has made use of the SIMBAD database, operated at CDS, Strasbourg, France and of the VizieR catalog access tool, CDS, Strasbourg, France. This publication makes use of data products from the Two Micron All Sky Survey, which is a joint project of the University of Massachusetts and the Infrared Processing and Analysis Center/California Institute of Technology, funded by the National Aeronautics and Space Administration and the National Science Foundation. This publication makes use of the AAVSO Photometric All-Sky Survey (APASS), funded by the Robert Martin Ayers Sciences Fund.
This research has benefited from the Ultracool RIZzo Spectral Library (http://dx.doi.org/10.5281/zenodo.11313), maintained by Jonathan Gagn\'e and Kelle Cruz. The authors would also like to thank our referee for excellent suggestions that improved the quality of this paper.

\emph{MED} wrote most of the manuscript, generated Figures~\ref{psf1219}, \ref{psf2250}, \ref{TS_optical}, \ref{opt_spec} and all Tables, obtained the CPAPIR data, led the companion identification analysis and wrote most of the computer codes. \emph{EA} generated Figures~\ref{sep1219}, \ref{pa_sep} and \ref{image2250}, led the CPAPIR data reduction, supervised the majority of the work and participated in the development of computer codes. \emph{JG} generated Figures~\ref{spectrecomp}, \ref{ellipsoid} and \ref{cmd}, led the
BANYAN~II analysis and the contamination analysis, obtained and reduced the IRTF/SpeX data, measured the FIRE radial velocities, convolved the ESPaDOnS spectra for template comparison and supervised the writing. \emph{RD} supervised the writing. \emph{LM} acquired and reduced the ESPaDOnS spectra and measured the ESPaDOnS radial velocities and $v\sin\,i$ and led the BANYAN analysis. \emph{JKF} acquired and reduced the FIRE spectra. \emph{DL} provided useful discussion on the development of the computer codes.

\facility{Magellan:Baade (FIRE); IRTF (SpeX); CFHT (ESPaDOnS); OMM (CPAPIR)}

\software{BANYAN \citep{2013ApJ...762...88M}, BANYAN~II \citep{2014ApJ...783..121G}, BANYAN~$\Sigma$ (Gagn\'e et al., submitted to ApJS), UPENA (Version 2.12)\footnote{ \url{http://www.cfht.hawaii.edu/Instruments/Upena/}} based on Libre-ESpRIT \citep{1997MNRAS.291..658D}, Firehose (Version 2.0) \citep{2009PASP..121.1409B,zenodofirehose} , spextool v4.0 beta IDL package \citep{2004PASP..116..362C}, xtellcor IDL routine \citep{2003PASP..115..389V}, Overleaf\footnote{\url{https://www.overleaf.com/}}}

\end{document}